\documentclass[letterpaper]{article} 
\usepackage{aaai23}  
\usepackage{times}  
\usepackage{helvet}  
\usepackage{courier}  
\usepackage[hyphens]{url}  
\usepackage{graphicx} 
\usepackage{tikz}
\usepackage{multicol}
\usepackage{fontawesome}
\usetikzlibrary{shapes, arrows.meta, positioning, calc, shadows}
\usepackage{array}
\usepackage{fontawesome}
\usepackage{natbib}  
\usepackage{caption} 
\usepackage{amsmath,amssymb}
\usepackage{booktabs}
\usepackage{algorithm}
\usepackage{algorithmic}

\usepackage{newfloat}
\usepackage{listings}

\title{Joint Learning of Policy with Unknown Temporal Constraints for Safe Reinforcement Learning}
\author {
    Lunet Yifru,\textsuperscript{\rm 1}
    Ali Baheri \textsuperscript{\rm 2}
}
\affiliations {
    \textsuperscript{\rm 1} West Virginia University\\
    \textsuperscript{\rm 2} Rochester Institute of Technology\\
    lay0005@mix.wvu.edu, akbeme@rit.edu
}

\usepackage{bibentry}

\begin{document}

\maketitle

\begin{abstract}
In many real-world applications, safety constraints for reinforcement learning (RL) algorithms are either unknown or not explicitly defined. We propose a framework that concurrently learns safety constraints and optimal RL policies in such environments, supported by theoretical guarantees. Our approach merges a logically-constrained RL algorithm with an evolutionary algorithm to synthesize signal temporal logic (STL) specifications. The framework is underpinned by theorems that establish the convergence of our joint learning process and provide error bounds between the discovered policy and the true optimal policy. We showcased our framework in grid-world environments, successfully identifying both acceptable safety constraints and RL policies while demonstrating the effectiveness of our theorems in practice.

\end{abstract}

\section{Introduction}

RL has emerged as a powerful computational approach for training agents to achieve complex objectives through interactions within stochastic environments \cite{sutton2018reinforcement}. RL algorithms have demonstrated significant success in a wide range of applications and domains \cite{singh2022reinforcement,razzaghi2022survey}. However, when deploying RL policies in real-world scenarios, particularly those involving safety-critical operations, ensuring the safety of the learning process becomes a paramount concern. Traditional RL algorithms tend to focus on reward maximization, which may inadvertently lead to violation of safety constraints. Safe RL aims to address this challenge by learning policies that not only maximize the expected return but also respect safety constraints throughout the learning process. One promising avenue of research in safe RL involves the use of formal methods, such as temporal logic, for specifying safety constraints in a mathematically rigorous manner. The use of temporal logic constraints in the reward function can enable RL agents to acquire policies that are not only efficient but also secure. However, this approach assumes the availability of accurate temporal logic specifications, which may not always be the case, especially in complex real-world environments. In this brief, we propose a novel framework for jointly learning RL policies and safety specifications. Our approach combines the strengths of RL for policy optimization with computational techniques for discovering temporal logic constraints from data. This joint learning strategy allows us to efficiently derive an optimal policy and a suitable safety constraint for a given environment, even in situations where the safety constraints are not explicitly provided in advance.

\section{Related Work}

\noindent{\textbf{Safe RL.}} Safe RL has garnered significant attention in recent years, as researchers aim to address safety concerns associated with deploying RL agents in safety-critical domains \cite{garcia2015comprehensive,gu2022review,baheri2020deep,baheri2022safe}. A prevalent approach to safe RL involves formulating the problem as a constrained optimization task, where the primary objective is to maximize the expected return while satisfying given safety constraints \cite{achiam2017constrained}. Another direction in safe RL is risk-sensitive RL, which aims to balance the trade-off between exploration, exploitation, and risk management \cite{mihatsch2002risk}. Risk-sensitive RL algorithms incorporate risk measures, such as conditional value-at-risk (CVaR) \cite{tamar2014policy} and risk envelope \cite{majumdar2017risk}, to guide the learning process. An additional approach to ensure safety in RL is through shielding, which intervenes in the agent's actions when it might violate safety constraints \cite{alshiekh2018safe}. Integrating formal methods, like temporal logic and Lyapunov-based techniques, into RL algorithms has emerged as a promising direction for safe RL \cite{hasanbeig2018logically,alur2023specification,chow2018lyapunov}.

\noindent{\textbf{STL Mining.}} STL has emerged as an essential formalism for specifying complex temporal properties and constraints in various applications, including robotics and cyber-physical systems. In recent years, researchers have focused on inferring or mining STL specifications from data, to facilitate the development of safe and robust systems. A key approach to mining STL from data is the use of algorithmic techniques, such as optimization-based algorithms \cite{abbas} and machine learning methods \cite{Fronda_2022}. Optimization-based techniques seek to minimize an objective function that captures the distance between the candidate STL formulas and the given data traces. Data-driven techniques have shown promise in learning STL specifications from data. Another direction in mining STL from data is the development of automated, scalable, and robust techniques for the discovery of interpretable STL specifications \cite{mohammadinejad2020interpretable}. \cite{BARTOCCI} provides a comprehensive survey of the various techniques for mining STL specifications from data. 


\section{Methodology} \label{sec:method}

We cast the joint learning of policy with unknown specifications as a bi-level optimization problem \cite{sinha2017review}. In this formulation, the upper level optimization aims to infer the correct STL safety constraint, while the lower level optimization focuses on learning the optimal policy under the inferred constraint. A human expert assists the learning by labeling trajectories based on their safety. In this context, safety is attained when a trajectory achieves the environmental objective without violating any safety constraints, i.e., the trace should have a positive robustness value against the true safety constraint. These components are iteratively called upon to simultaneously identify the optimal policy and the suitable STL constraint with the aid of the human expert. 
The outer loop, an evolutionary algorithm, is designed to infer both the template and the parameters of an STL specification that can classify the labeled dataset. This method is inspired by the work in \cite{ROGE}, which has been shown to result in simpler, more interpretable outputs, as well as an improved misclassification rate compared to those in \cite{BombaraandBelta, kong}. The algorithm implements the following procedures: random generation of the initial STL population, evaluation of fitness, $\mathcal{F}$, following the Eq. \ref{eqn:fitness}, ranking population members based on fitness, discarding the bottom half of the population, and applying genetic alterations such as mutation and crossover. For a positively labeled trace, $X_{p}$, and a negatively labeled trace, $X_{n}$, in their respective positive and negative datasets, $D_{p}$ and $D_{n}$, the fitness function is,

\begin{equation}
\mathcal{F}(\phi) = N_{\rho(\phi)\textsuperscript{+}\mid X_{p}}+ N_{\rho(\phi)\textsuperscript{-}\mid X_{n}} + \mid \overline{\rho}(\phi)_{D_{p}} - \overline{\rho}(\phi)_{D_{n}}\mid
\label{eqn:fitness}
\end{equation}
where, $\mathcal{F}$ is the fitness function for STL $\phi$. The first term in Eq. \ref{eqn:fitness} represents the number of true positive classifications for the positive samples, the second term represents the number of true negative classifications for the negative samples, and the third term computes the absolute value difference between the average of the robustness values, $\overline{\rho}(\phi)$, for samples in $D_{p}$ and $D_{n}$.


The inner loop is comprised of a logically-constrained Q-learning in which the reward is based solely on the robustness of a trajectory throughout an episode against a given STL constraint. The definition of the reward function is shown in Eq. \ref{eqn:reward}.
\begin{equation}
    \mathcal{R} = 
\begin{cases}
   \rho(\phi_{[0:T]}),& \text{if } \rho(\phi_{[0:T]}) < 0\\
    \rho(\phi_{[0:T]}) + 100,  & \text{if } \rho(\phi_{[0:T]}) \geq 0
\end{cases}
\label{eqn:reward}
\end{equation}
where, $\mathcal{R}$ is the reward value determined by the robustness degree $\rho(\phi)$ of the sample $s$ over the horizon of the STL, $T$. 

\begin{figure}[t]
  \centering
  \resizebox{\columnwidth}{!}{
\begin{tikzpicture}[
  node distance = 2cm,
  block/.style = {draw, rectangle, minimum width=3cm, minimum height=1.5cm, font=\footnotesize, rounded corners, top color=white, bottom color=blue!20, drop shadow},
  arrow/.style = {->, >={Stealth[length=3mm]}, line width=1mm}
]

\node[block] (EA) {STL Mining (Genetic Algorithm)};
\node[block, right=of EA] (Q) {Policy Learning (Q-learning)};
\node[block, below=of $(EA)!0.5!(Q)$, yshift=-1cm] (HE) {};
\node[font=\Large, anchor=center, yshift=3mm] at (HE) {\faUser};
\node[anchor=center, yshift=-3mm, font=\footnotesize] at (HE) {Human Expert};

\draw[arrow] (EA) -- (Q) node[midway, above, font=\footnotesize] {STL Constraint};
\draw[arrow] (Q) |- (HE) node[midway, right, font=\footnotesize] {Rollout Traces};
\draw[arrow] (HE) -| (EA) node[midway, left, font=\footnotesize] {Labeled Data};

\end{tikzpicture}
  } 
  \caption{Schematic representation of the integrated framework for concurrently learning STL constraints and optimal policies. The framework employs genetic algorithms for STL mining, Q-learning for policy learning, and incorporates human expert feedback for refining the learned constraints and policies.}
  \label{fig:framework}
\end{figure}
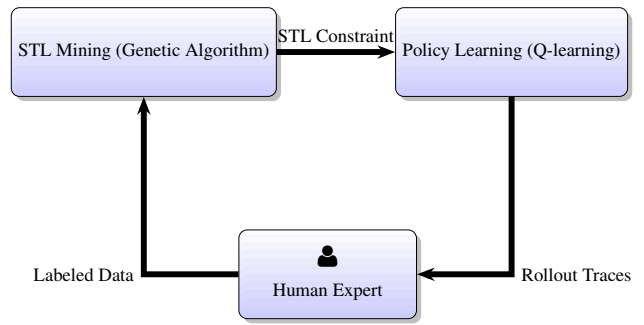

The reward is sparse because it is given at the end of an episode, and not at each step. This is due to the fact that, with timed STL constraints, the robustness value cannot be quantified at every step and can only be computed over a signal at least as long as the horizon, \textit{T}, of the STL. After training, the algorithm generates a certain number of rollout traces that are presented to the human expert for labeling based on their safety, which is our final component. In our experiments, we have automated the human labeling process by computing the robustness value of the traces against the true safety constraint of the environment. However, it is important to note that this is only done for automation purposes, and as per the basis of our problem, this true safety constraint is unknown, and traces should actually be labeled by an expert. The labeled traces are then used as input to the evolutionary algorithm. This process is repeated iteratively until convergence, which, in this case, is defined by the number of rollout traces that are labeled safe by the human expert. This metric is chosen because the safety of the rollout traces reflects the quality of the STL used as a safety constraint as well as the quality of the policy generated using that constraint. The framework is depicted graphically in Fig. \ref{fig:framework}.


\vspace{-4 mm}
\section{Results}
We consider the problem of implementing RL algorithms in an environment where the safety constraint is unknown in advance. Specifically, our goal is to simultaneously infer the correct STL safety constraint and an optimal policy. To evaluate our framework, we have designed grid-world environments of varying sizes for an agent to navigate through to reach a goal at a specific location, under temporal constraints. Initially, neither the goal location nor the time constraint are known, making it impossible to design a traditional reward function. The problem was initiated with $1000$ random $2$-dimensional coordinate traces within the environment, which were then labeled by a human to create a dataset for the evolutionary algorithm. The algorithm proceeds with the steps outlined in the methodology until the number of safe traces, as labeled by the human expert, meets a certain threshold. The experiment was performed on \text{6$\times$6}, \text{8$\times$8}, and \text{10$\times$10} grid environments. The outputs were evaluated based on two metrics: (i) the change in the percentage of the number of unsafe traces from the first batch of rollout traces to the last batch and (ii) the average misclassification rate (MCR) of the inferred STL specification against a dataset labeled by the human expert. The first metric evaluates how closely the inferred STL specification is able to capture the true environmental constraints by assessing how the number of unsafe samples in the rollout traces has decreased over the iterations, indicating the STL specification is getting closer to the true (but unknown) constraint. The second metric conveys the classification capability of the inferred STL against labeled datasets. It quantifies how well the STL distinguishes between safe and unsafe trajectories, as compared to that of a human expert. The results are given in Table \ref{results table}.


\begin{table}[t]
\centering
\caption{Percentage of unsafe traces (per rollout sample) at the beginning and the end of the process, and average MCR for inferred STL.}
\label{results table}
\begin{tabular}{l|c|c|c}
\toprule
Size & \multicolumn{2}{c|}{$\%$ of Unsafe Traces} & Inferred STL \\
\cline{2-3}
& First rollout & Last rollout & MCR \\
\midrule
6$\times$6  & 73.2$\%$ & 1.2$\%$ & 0.02$\pm$0.0013 \\
8$\times$8  & 86.7$\%$ & 4.3$\%$ & 0.04$\pm$0.001 \\
10$\times$10 & 91.4$\%$ & 11.1$\%$ & 0.06$\pm$0.009 \\
\bottomrule
\end{tabular}
\end{table}

\section{Theoretical Results}

To complement these empirical findings, we now delve into the theoretical underpinnings of our approach. In this section, we will present the theoretical results that support the convergence properties and error bounds of our joint learning framework, providing a more rigorous understanding of its performance.

\subsection{Joint Convergence of the Inner Loop and Outer Loop}

Our objective is to demonstrate that the combined convergence of the inner and outer loops results in the overall convergence of the framework. In essence, we aim to prove that when the inner loop (Q-learning) reaches an optimal policy and the outer loop (evolutionary algorithm for STL synthesis) attains an optimal STL constraint, the entire framework converges to a stable solution. By establishing these two implications, we can show that the proposed framework effectively tackles the given problem and converges to a solution that satisfies both the policy's optimality and the STL constraint's optimality. To prove this joint convergence, we must address the following two implications:

\begin{itemize}
    \item Convergence of the inner loop (Q-learning) to an optimal policy $\pi^{\ast}$ implies the convergence of the outer loop (evolutionary algorithm) to an optimal STL constraint $\phi^{\ast}$.

    \item Convergence of the outer loop (evolutionary algorithm) to an optimal STL constraint $\phi^{\ast}$ implies the convergence of the inner loop (Q-learning) to an optimal policy $\pi^{\ast}$.

\end{itemize}

\noindent{\textbf{Implication 1.}} This implication aims to show that if the inner loop converges to an optimal policy, the outer loop converges to an optimal STL constraint. This is achieved by demonstrating that the fitness function is maximized for the optimal STL constraint and that the evolutionary algorithm converges to this optimal constraint under certain conditions. Assume that the inner loop converges to an optimal policy $\pi^{\ast}$. Under this assumption, we need to prove that the outer loop converges to an optimal STL constraint $\phi^{\ast}$. We approach this by showing that:

\begin{itemize}

\item The fitness function $\mathcal{F}(\pi^{\ast}, \phi)$ is maximized for the optimal STL constraint $\pi^{\ast}$. This can be done by analyzing the properties of the fitness function and the reward function $\mathcal{R}(\pi,\phi)$ under the optimal policy.

\item  The evolutionary algorithm for STL synthesis converges to the optimal STL constraint $\pi^{\ast}$ under certain conditions, such as proper selection pressure, sufficient exploration, and well-defined mutation and crossover operators.

\end{itemize}
By showing that the fitness function is maximized for the optimal STL constraint, and that the evolutionary algorithm converges to this optimal constraint, we establish the convergence of the outer loop under the assumption that the inner loop converges to the optimal policy.

\noindent{\textbf{Implication 2.}} This implication aims to show that if the outer loop converges to an optimal STL constraint, the inner loop converges to an optimal policy. This is achieved by demonstrating that the reward function provides the necessary guidance under the optimal STL constraint, and that the Q-learning algorithm converges to the optimal policy under this guidance. We assume that the outer loop converges to an optimal STL constraint $\pi^{\ast}$. Under this assumption, we need to prove that the inner loop converges to an optimal policy $\pi^{\ast}$. We approach this by showing that: 

\begin{itemize}
    \item The reward function $\mathcal{R}(\pi, \phi^{\ast})$ has the necessary properties to guide the Q-learning algorithm towards the optimal policy. This can be done by analyzing the reward function under the optimal STL constraint and ensuring that it provides proper guidance and exploration-exploitation trade-off.

    \item The Q-learning algorithm converges to the optimal policy $\pi^{\ast}$ under certain conditions, such as proper learning rates, sufficient exploration, and well-defined state and action spaces.
\end{itemize}
We demonstrate the convergence of the inner loop by illustrating that the reward function offers the needed guidance when operating under the optimal STL constraint, and that the Q-learning algorithm converges to the optimal policy with this guidance. This convergence is established based on the assumption that the outer loop successfully converges to the optimal STL constraint.

\subsubsection{PROOF of Implication 1.} \noindent{\textit{(a) Maximizing the fitness function for the optimal STL constraint.}}

Here, we provide a mathematical presentation of Implication 1. Let us first define the key components of the framework: (i) policy: $\pi : \mathcal{S} \to \mathcal{A}$, a mapping from states to actions, (ii) reward function: $\mathcal{R}(\pi, \varphi) : \Pi \times \Phi \to \mathbb{R}$, a function that measures the reward for a given policy $\pi$ and STL constraint $\varphi$, (iii) fitness function: $\mathcal{F}(\pi, \varphi) : \Pi \times \Phi \to \mathbb{R}$, a function that measures the fitness of a given policy $\pi$ and STL constraint $\varphi$. Now, let's proceed with the proof of Implication 1:

Assume that the inner loop converges to an optimal policy $\pi^\ast$. We want to show that the fitness function $\mathcal{F}(\pi^{\ast}, \varphi)$ is maximized for the optimal STL constraint $\varphi^\ast$. Let $\mathcal{R}^\ast = \max_{\pi, \varphi}\mathcal{R}(\pi, \varphi)$ be the maximum achievable reward. We know that the reward function $\mathcal{R}(\pi, \varphi)$ is maximized for the optimal policy $\pi^\ast$ and the optimal STL constraint $\varphi^\ast$, i.e., $\mathcal{R}(\pi^{\ast}, \varphi^{\ast}) = \mathcal{R}^*$. We define the fitness function $\mathcal{F}(\pi, \varphi)$ as a function of the reward function $\mathcal{R}(\pi, \varphi)$:

$$\mathcal{F}(\pi, \varphi)=\frac{R(\pi, \varphi)}{R^*}$$
since $\mathcal{R}(\pi^\ast, \varphi^\ast) = \mathcal{R}^*$, we have:

$$\mathcal{F}\left(\pi^*, \varphi^*\right)=\frac{\mathcal{R}\left(\pi^*, \varphi^*\right)}{\mathcal{R}^*}=\frac{\mathcal{R}^*}{\mathcal{R}^*}=1$$
This result shows that the fitness function $\mathcal{F}(\pi^\ast, \varphi)$ is indeed maximized for the optimal STL constraint $\varphi^\ast$, given that the inner loop converges to the optimal policy $\pi^\ast$. The maximum value of the fitness function is 1, which occurs when both the policy and the STL constraint are optimal.

\noindent{\textit{(b) Convergence of the evolutionary algorithm for STL synthesis.}} 

Now, we provide insights into the convergence of the evolutionary algorithm for STL synthesis. Let $\varphi_i$ be the STL constraint at iteration $i$ of the evolutionary algorithm. We want to show that the evolutionary algorithm converges to the optimal STL constraint $\varphi^*$ under certain conditions. Let $P(\varphi_i)$ be the probability distribution of the STL constraint population at iteration $i$. The evolutionary algorithm updates $P(\varphi_i)$ through selection, mutation, and crossover operators. Let $P_{\text{sel}}(\varphi_i)$, $P_{\text{mut}}(\varphi_i)$, and $P_{\text{cross}}(\varphi_i)$ be the updated probability distributions after applying the selection, mutation, and crossover operators, respectively. Then, the probability distribution at iteration $i+1$ is given by:

$$P\left(\varphi_{i+1}\right)=P_{\text {cross }}\left(P_{\text {mut }}\left(P_{\text {sel }}\left(\varphi_i\right)\right)\right)$$
Under proper selection pressure, sufficient exploration, and well-defined mutation and crossover operators, it can be shown that the evolutionary algorithm converges to the optimal STL constraint $\varphi^*$ as the number of iterations approaches infinity:

$$\lim _{i \rightarrow \infty} P\left(\varphi_i\right)=\delta\left(\varphi-\varphi^*\right)$$
where $\delta(\cdot)$ is the Dirac delta function, meaning that the probability distribution converges to a distribution concentrated on the optimal STL constraint $\varphi^\ast$. Through the demonstration of both points (a) and (b), we confirm that the outer loop approaches the optimal STL constraint, denoted as $\varphi^\ast$, provided that the inner loop approaches the optimal policy, denoted as $\pi^\ast$. This evidence indicates that the combined convergence of both the inner and outer loops results in the comprehensive convergence of the entire framework.

\subsection{PROOF of Implication 2.} Assume that the outer loop converges to an optimal STL constraint $\varphi^\ast$. Under this assumption, we need to prove that the inner loop converges to an optimal policy $\pi^\ast$. We approach this by showing that:

\begin{itemize}
    \item The reward function $\mathcal{R}(\pi, \varphi^\ast)$ has the necessary properties to guide the Q-learning algorithm towards the optimal policy. This can be done by analyzing the reward function under the optimal STL constraint and ensuring that it provides proper guidance and exploration-exploitation trade-off. Specifically, we show that $\mathcal{R}(\pi, \varphi^\ast)$ is Lipschitz continuous and has a unique maximum at the optimal policy $\pi^*$. Moreover, the reward function should encourage sufficient exploration of the state-action space while exploiting the knowledge acquired during the learning process.

    \item The Q-learning algorithm converges to the optimal policy $\pi^\ast$ under certain conditions, such as proper learning rates, sufficient exploration, and well-defined state and action spaces. According to the Q-learning convergence theorem, the Q-learning algorithm converges to the optimal Q-function $Q^\ast(s,a)$ if the following conditions are satisfied:
\end{itemize}

\begin{enumerate}
    \item Each state-action pair $(s, a)$ is visited infinitely often, i.e., $\lim_{t\to\infty} N_t(s, a) = \infty$, where $N_t(s, a)$ is the number of visits to the state-action pair $(s, a)$ up to time $t$.

     \item The learning rate $\alpha_t(s, a)$ satisfies $\sum_{t=1}^{\infty} \alpha_t(s, a) = \infty$ and $\sum_{t=1}^{\infty} \alpha_t^2(s, a) < \infty$. This condition ensures that the learning rate decays slowly enough to guarantee convergence.
     
\end{enumerate}
Assuming that we have the optimal STL constraint $\varphi^\ast$, we consider well-defined state and action spaces, along with an exploration strategy (such as an $\epsilon$-greedy approach) that ensures each state-action pair is visited infinitely often. If these conditions hold, the Q-learning algorithm converges to the optimal policy $\pi^\ast$. The convergence of the inner loop is based on two critical observations: first, the reward function provides essential guidance when used under the optimal STL constraint, and second, the Q-learning algorithm moves towards the optimal policy when steered by this guidance. This convergence is contingent on the outer loop effectively converging to the optimal STL constraint.

\subsection{Bounds on the Error}

Deriving bounds on the error between the discovered policy and the true optimal policy involves analyzing the mathematical relationship between the error and various factors influencing it. Here's a possible way to approach this analysis: Let $\pi^*$ be the true optimal policy and $\pi'$ be the discovered policy. Define the error between these policies as:

$$\epsilon\left(\pi^{\prime}, \pi^*\right)=E\left[R\left(s, \pi^*(s)\right)-R\left(s, \pi^{\prime}(s)\right)\right]$$
where $E[\cdot]$ denotes the expected value, $\mathcal{R} (s, a)$ is the reward function for taking action $a$ in state $s$, and the expectation is taken over all states $s$ in the state space. Now, consider the following factors that may affect the error bounds:

\begin{itemize}
    \item Granularity of the state abstraction (denoted by $G$): A coarse state abstraction may lead to a larger error between the discovered policy and the true optimal policy. The impact of state abstraction granularity on the error can be represented as: $$\epsilon_G(G) \leq C_1 \cdot G$$ where $C_1$ is a constant that depends on the problem's specific characteristics.

    \item Quality of the learned STL specifications (denoted by $Q$): If the learned STL specifications are not accurate or expressive enough, the error between the discovered policy and the true optimal policy may be larger. The impact of the quality of the learned STL specifications on the error can be represented as: $$\epsilon_Q(Q) \leq C_2 \cdot(1-Q)$$ where $C_2$ is a constant that depends on the problem's specific characteristics.

    \item Amount of human feedback provided (denoted by $H$): Human feedback can help guide the learning process and reduce the error between the discovered policy and the true optimal policy. The impact of the amount of human feedback on the error can be represented as: $$\epsilon_H(H) \leq C_3 \cdot e^{-H}$$ where $C_3$ is a constant that depends on the problem's specific characteristics, and $e^{-H}$ represents the exponential decay in error with increasing human feedback.

\end{itemize}
Integrating these individual error bounds, we obtain an overall error bound:

$$\epsilon\left(\pi^{\prime}, \pi^*\right) \leq C_1 \cdot G+C_2 \cdot(1-Q)+C_3 \cdot e^{-H} .$$
The error bound reveals how the error is influenced by various contributing factors, including the level of detail in state abstraction, the accuracy of the learned STL specifications, and the extent of human feedback received. By carefully analyzing the error bound, we gain insight into the trade-offs between these factors. Armed with this knowledge, we can formulate strategies that minimize the error and ultimately improve the effectiveness of the bi-level optimization framework.

\section{Conclusions}

In this paper, we have studied a joint learning framework for the safety constraint and the RL policy of an environment where the safety constraints are unknown \emph{a priori}. We have implemented an algorithm that optimizes the safety constraint and the RL policy simultaneously. Our preliminary results have shown that our framework is capable of identifying safety constraints that are suitable for the environment and an optimal RL policy that results in safe behavior. Future directions for this work will include testing our algorithm in complex environments, assessing adaptability, and improving the algorithm's computational efficiency.

\bibliography{aaai23}


\end{document}